\documentclass[sort&compress,number]{elsarticle}
\usepackage[utf8]{inputenc}
\usepackage{amsmath}
\usepackage{amssymb}
\usepackage{amsthm}
\usepackage{algorithm2e}
\usepackage{setspace}
\usepackage{comment}
\newtheorem{prop}{Proposition}[section]
\newtheorem{definition}{Definition}[section]
\newtheorem{lemma}{Lemma}[section]
\journal{arXiv.org}
\title{Counting and Verifying Abelian Border Arrays of Binary Words}
\author[buet]{Mursalin Habib}
\ead{mursalinmail@gmail.com}
\author[buet]{Md. Salman Shamil}
\ead{1505021.mss@ugrad.cse.buet.ac.bd}
\author[buet]{M. Sohel Rahman\corref{cor1}}
\ead{msrahman@cse.buet.ac.bd}
\address[buet]{A\(\ell\)EDA Group, Department of CSE, BUET, Dhaka-1000, Bangladesh}
\cortext[cor1]{Corresponding author: \ead{msrahman@cse.buet.ac.bd}}

\date{}

\begin{document}

\begin{abstract}
    In this note, we consider the problem of counting and verifying abelian border arrays of binary words. We show that the number of valid abelian border arrays of length \(n\) is \(2^{n-1}\). We also show that verifying whether a given array is the abelian border array of some binary word reduces to computing the abelian border array of a specific binary word. 
    Thus, assuming the word-RAM model, we present an \(O\left(\frac{n^2}{\log^2n}\right)\) time algorithm for the abelian border array verification problem.
    
    
\end{abstract}

\maketitle
\section{Introduction}
In recent years, there has been much interest in the field of abelian stringology. The central concept of abelian stringology is that of \textit{abelian equivalence}: two strings are abelian equivalent if they have the same letters with the same multiplicities. For example, the words \texttt{LISTEN} and \texttt{SILENT} are abelian equivalent. We refer the reader to Section \ref{preliminaries} for more precise definitions, especially in the case of binary words. 

By substituting string equality with abelian equivalence, we can get abelian analogs of many natural string problems and regularities, e.g., abelian pattern matching \cite{liptak,rahman,Moosa-permutation}, abelian borders \cite{abelianborder}, abelian squares \cite{cummings}, common abelian factors \cite{costas} - just to name a few of the topics touched on in recent literature.

There has also been a long practice of studying string inference or string reverse engineering problems where, given an instance of a string data structure, one attempts to find a string that generates that given data structure (or report if none exists). The first string reverse engineering problem was introduced by Fran\u ek \textit{et al.} \cite{franek} who proposed a method to check if any integer array was the border array of some string. Since then a plethora of string inference problems have been studied in the literature (e.g., \cite{abelianborder,Infer-Cover,Infer-Baisya-CArray, Infer-Indeter,Infer-Daykin,Infer-Helling,NazeenRR12,BaisyaFR13,MoosaNRR13,AlatabbiRS15}).

In this note, we study the abelian analog of the problem introduced in \cite{franek} for binary words. In particular, given an integer array, we propose a method that decides whether the array is the abelian border array of some binary words. We show that this ``abelian border array verification" problem reduces to computing the abelian border array of a specific binary word (Section \ref{sec:verify}). In addition, we count the number of ``valid" abelian border arrays (Section \ref{sec:count}) and present some properties thereof (Section \ref{sec:verify}). We also briefly discuss the problems for larger alphabets (Section \ref{sec:disc}).  

\section{Preliminaries}
\label{preliminaries}

Let \(\Sigma\) be a finite set of \textit{letters} called an \textit{alphabet}. Then \(\Sigma^*\) is the set of all finite words over \(\Sigma\). For a binary alphabet, we assume \(\Sigma = \{0,1\}\) and \(w\) is called a \textit{binary word} if \(w\in \Sigma^*\). The \textit{length} of a word \(w\) is denoted by \(|w|\). We will denote by \(\Sigma^n\) the set of all words of length \(n\) over \(\Sigma\). The word \(w\) will often be represented as \(w[1]w[2]\cdots w[|w|]\) where \(w[i]\) refers to the \(i\)-th letter of \(w\). For \(1\leq i \leq j \leq |w|\), we let \(w[i\cdots j]\) denote the \(j-i+1\) length word \(w[i]w[i+1]\cdots w[j]\), also referred to as a factor of $w$. Furthermore, $w[i\cdots j]$ is called a prefix (suffix) if $i=1$ ($j = |w|$).
A prefix or a suffix of \(w\) is called \textit{proper} if it is not equal to \(w\). Given a binary word \(w\), let \(ones(w)\) be equal to the number of \(1\)'s in \(w\). Although not very common, we sometimes conveniently use the following notation: if $w = w[1]w[2]\cdots w[|w|-1]w[|w|]$, $w' = w[1]w[2]\cdots w[|w|-1]]$ and $w[|w|] = \alpha$, then $w=w'\alpha.$

Two binary words \(x\) and \(y\) of equal length are said to be \textit{abelian equivalent} if \(ones(x)=ones(y)\). An \textit{abelian border} of a binary word \(w\) is a proper prefix of \(w\) that is abelian equivalent to a proper suffix of \(w\). Our results center around a data structure called the abelian border array which we define below.

\begin{definition}
    Let \(x\) be a binary word. The abelian border array of \(x\), denoted by \(\pi_x\), is an array of length \(|x|\) such that for \(1\leq i\leq |x|\), \(\pi_x[i]\) contains the length of the longest abelian border of \(x[1\cdots i]\). An array \(\pi\) is called a valid abelian border array if \(\pi=\pi_x\) for some binary word \(x\); in that case, \(x\) is called a generating word of \(\pi\).
\end{definition}

For example, \(\pi_{0001001}= \left(0, 1, 2, 0, 4, 5, 0\right)\). Also, \(\left(0, 1, 2, 0, 4, 4\right)\) is not a valid abelian border array since there does not exist any binary word \(x\) such that \(\pi_x= \left(0, 1, 2, 0, 4, 4\right)\).

Next, we introduce the notion of abelian border equivalence. Two binary words \(x\) and \(y\) are called \textit{abelian border equivalent} if \(\pi_x=\pi_y\), i.e., they have the same abelian border array. For example, \(\pi_{0100}=\pi_{1011}\) and hence the two strings, \(0100\) and \(1011\) are abelian border equivalent.

Finally, it comes in handy to define the complement of a binary word.

\begin{definition}
\label{def:comp}
    Given a binary word \(x\), the complement of \(x\), denoted by \(\overline{x}\), is a binary word of length \(|x|\) such that for \(1\leq i\leq |x|\)

    \[\overline{x}[i] = \begin{cases}
      1 & \text{if } x[i]=0 \\
      0 & \text{otherwise.}
   \end{cases}
\]
\end{definition}

\section{Results}
\label{results}

\subsection{Counting the Number of Valid Abelian Border Arrays}\label{sec:count}

The first problem we tackle is counting the number of valid abelian border arrays of a particular length. More formally, suppose that \(\Pi_n\) is the set of \(n\)-length arrays such that for each \(\pi \in \Pi_n\) there exists a binary word \(x\) with \(|x|=n\) and \(\pi=\pi_x\). Let \(T_n\) be the number of arrays in the set \(\Pi_n\), i.e., \(T_n = |\Pi_n|\).
We prove the following proposition.
\begin{prop}\label{prop_count}
\(T_n=2^{n-1}\).
\end{prop}

To prove Proposition \ref{prop_count}, we first prove the following lemma.
\begin{lemma}
    If \(x\) and \(y\) are two different binary words that are abelian border equivalent, then \(y=\overline{x}\).
\end{lemma}

\begin{proof}
    We prove this by induction on the length of \(x\) and \(y\). Clearly, the claim is true when \(|x|=|y|=1\). Assume the claim is true whenever \(|x|=|y|=n-1\).

    Now let \(x\) and \(y\) be two different abelian border equivalent binary words of length \(n>1\). If we have \(x[1\cdots n-1] = y[1\cdots n-1]\), then the fact that \(x\) and \(y\) are different will force \(\pi_x\neq \pi_y\) making \(x\) and \(y\) abelian border non-equivalent, contradicting our hypothesis.

    So, \(x[1\cdots n-1]\neq y[1\cdots n-1]\). Since \(\pi_{x[1\cdots n-1]}=\pi_{y[1\cdots n-1]}\), by the inductive hypothesis, we have \(y[1\cdots n-1]=\overline{x[1\cdots n-1]}\). Therefore, it suffices to show that \(x[n]\neq y[n]\).

    For the sake of contradiction, let \(x[n]=y[n]=\alpha\). Since \(y[1\cdots n-1]=\overline{x[1\cdots n-1]}\), it follows that \(x[1]\neq y[1]\). Without loss of generality, we can assume \(x[1]=\alpha\). However, this forces \(\pi_x[n]=n-1\) and \(\pi_y[n]<n-1\) contradicting the assumption that \(x\) and \(y\) are abelian border equivalent.
\end{proof}

From the lemma above, it is clear that every valid abelian border array has exactly two generating words and they are complements of each other. In other words, the generating word of a valid abelian border array is unique \textit{up to complementation}. As a result, as far as abelian border arrays of binary words are concerned, it suffices to only consider words that start with a \(0\). We make this notion formal by defining \textit{the} generating word of a valid abelian border array to be the generating word that starts with a \(0\).

The key to finding \(T_n\) lies in the observation that given a valid abelian border array of length \(n-1\), there are exactly two ways to extend it into a valid abelian border array of length \(n\). The following lemmas explore this idea.

\begin{lemma}\label{lemma:x0}
Let \(\pi\) be a valid abelian border array of length \(n-1\) and \(x\) be the generating word thereof. Then \(\pi_{x0}[n]=n-1\).
\end{lemma}
\begin{proof}
Consider the word \(y = x0\). By definition, we have \(|y|=n\) and \(y[1] = x[1] = 0 = y[n]\). So, \(ones(y[1\cdots n-1]) = ones(x[2\cdots n-1]) = ones(y[2\cdots n])\). Therefore \(y = x0\) has an abelian border of length \(n-1\) and the result follows.
\end{proof}

\begin{lemma}
    \label{recurrenceLemma}
    Let \(\pi\) be a valid abelian border array of length \(n-1\). If \(S\) is the set of all possible non-negative integers \(k\) such that appending \(k\) to the end of \(\pi\) gives a valid abelian border array of length \(n\), then \(|S|=2\).
\end{lemma}

\begin{proof}
    Let \(x\) be the generating word of \(\pi\). Now consider the string $x0$. 
    By Lemma \ref{lemma:x0}, we must have \(\pi_{x0}[n]=n-1\). Therefore, \(n-1\in S\). The other element in \(S\) is also completely determined by \(x\). In fact it is equal to \(\pi_{x1}[n]\). The fact that there are no other elements in \(S\) follows from the uniqueness of \(x\).
\end{proof}

For example, \(\left(0, 0, 0, 3, 3\right)\) is a valid abelian border array for which the generating word is \(01101\) and in this case, \(S=\{3, 5\}\). Note that \(\left(0, 0, 0, 3, 3, 3\right)=\pi_{011011}\) and \(\left(0, 0, 0, 3, 3, 5\right)=\pi_{011010}\).

We are now ready to prove the main result of this section.


\begin{proof}[Proof of Proposition \ref{prop_count}]
    Since prefixes of valid abelian border arrays are themselves valid abelian border arrays, the only way to get a valid abelian border array of length \(n\) is to extend a valid abelian border array of length \(n-1\). From Lemma \ref{recurrenceLemma}, it follows that \(T_n=2T_{n-1}\). Since \(T_1=1\), we have \(T_n=2^{n-1}\).
\end{proof}

\subsection{Verifying Valid Abelian Border Arrays}\label{sec:verify}

Now we turn to the problem of verifying abelian border arrays. More formally, given an array \(\pi\), we want to find whether or not it is a valid abelian border array. In addition, if the answer is positive, we want to find the generating word thereof. We first look at some general properties of abelian border arrays.
%

\begin{prop}
Let \(x\) be a binary word of length \(n\). For \(1\leq i \leq n\), the length of the shortest non-empty abelian border of \(x[1\cdots i]\) is equal to \(i-\pi_x[i]\) provided that \(\pi_x[i]\neq 0\).
\end{prop}

\begin{proof}
\label{longshort}
This follows from the fact that if a word \(w\) has an abelian border of length \(i\), then it also has an abelian border of length \(|w|-i\). It is then immediately clear why the lengths of the longest and shortest non-empty abelian borders should be related in this way.
\end{proof}

\begin{prop}
\label{i-1}
Let \(\pi\) be a valid abelian border array of length \(n\) and let \(x\) be the generating word of \(\pi\). For \(1\leq i\leq n\), \(\pi[i]=i-1\) if and only if \(x[i]=0\).
\end{prop}

\begin{proof}
The claim is obviously true if \(i=1\). So, let \(i>1\). Since \(x\) is the generating word of \(\pi\), \(x[1]=0\). By proposition \ref{longshort}, \(\pi[i]=i-1\) if and only if the length of the shortest non-empty abelian border of \(x[1\cdots i]\) is \(1\). But since \(x[1]=0\), this can happen if and only if \(x[i]=0\).
\end{proof}

\begin{prop}
\label{monotonicity}
Let \(x\) be a binary word of length \(n\) such that \(x[1]=0\). For \(1 < i \leq n\), if \(x[i]=1\), then \(\pi_x[i]\leq \pi_x[i-1]\).
\end{prop}

\begin{proof}

Let \(k=i-1-\pi_x[i-1]\) and \(k'=i-\pi_x[i]\). By proposition \ref{longshort}, \(k\) and \(k'\) are the lengths of the shortest non-empty abelian borders of \(x[1\cdots i-1]\) and \(x[1\cdots i]\) respectively. Therefore, it suffices to show that \(k'\geq k\).

Since \(k\) is the length of the shortest non-empty abelian border of \(x[1\cdots i-1]\), for \(1\leq j < k\), \(ones(x[1\cdots j]) < ones(x[i-j\cdots i-1])\). However, since \(x[i]=1\), \(ones(x[i-j+1\cdots i])\geq ones(x[i-j \cdots i-1])\). So, \(ones(x[1\cdots j]) < ones(x[i-j+1\cdots i])\) for \(1\leq j < k\). So, we can conclude that for \(1\leq j < k\), we must have \(ones(x[1\cdots j]) \neq ones(x[i-j+1\cdots i])\). Therefore, \(k'\) can not be smaller than \(k\).

\end{proof}

Propositions \ref{i-1} and \ref{monotonicity} provide us with an insight into the structure of valid abelian border arrays. They tell us that consecutive elements of a valid abelian array can not increase ``slowly". Given a binary word \(x\) with \(x[1]=0\), \(\pi_x[i]\) either jumps up to \(i-1\) (happens when \(x[i]=0\)) or stays at most as high as \(\pi_x[i-1]\) (happens when \(x[i]=1\)). A long run of 1s in \(x\) eventually brings \(\pi_x[i]\) down to \(0\); after which a \(0\) in \(x\) brings it again up to \(i-1\). 
 
Proposition \ref{i-1} actually suggests a direct algorithm for our verification problem as we show below.

\begin{prop}\label{prop:suff}

Let \(\pi\) be an array of length \(n\). We define \(x_\pi\) to be a binary word of length \(n\) such that

\[x_\pi[i] = \begin{cases}
      0 & \text{if } \pi[i]=i-1 \\
      1 & \text{otherwise.}
   \end{cases}
\]

If \(\pi\) is a valid abelian border array, then \(\pi=\pi_{x{_\pi}}\).\qed

\end{prop}

So, the problem of checking whether an array \(\pi\) is a valid abelian border array reduces to computing the abelian border array of a specific binary word \(x_{\pi}\). If the computed abelian border array matches \(\pi\), we output \textbf{yes} along with the word \(x_\pi\). Otherwise, we output \textbf{no}.

The abelian border array of a binary word can be computed naively in \(O(n^2)\) where \(n\) is the length of the word. But a recent result by Kociumaka \textit{et al}. \cite{kociumaka} shows that it can be done in \(O\left(\frac{n^2}{\log^2n}\right)\) time assuming the word-RAM model.

\begin{prop}
Assuming the word-RAM model, the valid abelian border array verification problem can be solved in \(O\left(\frac{n^2}{\log^2n}\right)\) time.\qed 
\end{prop}

\section{Extending to Larger Alphabets}

A natural thing to do is to try extending these results for words over larger alphabets. However, the problem becomes quickly difficult even for ternary words. The main reason is that it is hard to find a good characterization of abelian border equivalent words on larger alphabets. Two words can be very different but can still give the same abelian border array. For example, the words 011021 and 012022 are abelian border equivalent but at a first glance, they do not look anything alike.

Despite this, it is possible to come up with upper bounds for the answer to the counting problem for larger alphabets. The key idea is the following definition.

\begin{definition}
    Two words \(w_1\) and \(w_2\) with \(|w_1|=|w_2|=n\) are said to be letter-equivalent if for all \(1\leq i, j \leq n\), \(w_1[i]=w_1[j]\) if and only if \(w_2[i]=w_2[j]\).
\end{definition}

Note that letter-equivalent words are a generalization of complement words (Definition \ref{def:comp}) for larger alphabets. Clearly, if two words are letter-equivalent, then they are abelian border equivalent. However, the converse is not necessarily true for words on larger alphabets. We have already provided an example of this: the two words 011021 and 012022, despite not being letter-equivalent, are abelian border equivalent.

Letter-equivalence, as the name suggests, is an equivalence relation on the set \(\Sigma^n\) of all words of length \(n\) over \(\Sigma\). Therefore, the set of distinct equivalence classes of letter-equivalence forms a partition of \(\Sigma^n\). Clearly, the number of parts in this partition is an upper bound for \(T_n\). Thus we have the following two results.

\begin{prop}
Let \(T_n\) be the number of \(n\)-length arrays \(\pi\) such that there exists a word \(x\) over \(\Sigma=\{0, 1, 2\}\) with \(\pi=\pi_x\). Then \(T_n\leq \frac{3^{n-1}+1}{2}\). 
\end{prop}

\begin{proof}
We count the number of distinct equivalence classes of letter-equivalence in \(\Sigma^n\). Out of the \(3^n\) words that form \(\Sigma^n\), the \(3\) words that contain only one letter are in an equivalence class of their own. Each of the remaining \(3^n-3\) words are in an equivalence class with \(5\) other words that can be found by simply relabeling the letters (as an example, the word \(0110\) is in an equivalence class with the five words \(0220, 1001, 1221, 2002,\) and \(2112\)). Therefore, \(T_n \leq 1+\frac{3^n-3}{6}=\frac{3^{n-1}+1}{2}\).
\end{proof}

\begin{prop}
Let \(n\geq 2\) be an integer and \(\Sigma=\{0, 1, 2, \cdots, n-1\}\). If \(T_n\) is the number of \(n\)-length arrays \(\pi\) such that there exists a word \(x\) over \(\Sigma\) with \(\pi=\pi_x\), then \(T_n \leq B_n\) where \(B_n\) is the \(n\)th Bell number.
\end{prop}

\begin{proof}
Each word \(w\in \Sigma^n\) induces a partition of the indices \(1, 2, 3, \cdots , n\) in the following way: for all \(1\leq i < j \leq n\), the indices \(i\) and \(j\) are in the same part of the partition if and only if \(w[i]=w[j]\). Two words \(w_1, w_2 \in \Sigma^n\) are letter-equivalent if and only if they induce the same partition of the indices. Therefore, an upper bound on \(T_n\) is the number of ways you can partition the set \(\{1, 2, 3, \cdots , n\}\). This number is precisely \(B_n\) \cite{gardner}.
\end{proof}

Therefore, for an unbounded alphabet \(T_n\) is upper-bounded by the \(n\)th Bell number. However, this bound is very loose and does not offer much insight into the structure of valid abelian border arrays. 

\section{Conclusion}\label{sec:disc}

In this note, we have taken on the problem of inferring a binary word from its abelian border array. Although regular string inference problems are abundant in the literature, inference problems of the abelian variety are surprisingly rare. We hope our work will be one of the first of many ventures into the word of abelian string inference problems. 

Possible future work might include extending our results for words over larger alphabets. However, as the last section shows, doing this is non-trivial. Another line of work would be to ask if it is actually \textit{necessary} to compute abelian border arrays at all to solve the verification problem. We have shown that it is sufficient (Proposition \ref{prop:suff}). But it might be possible for some other verification algorithm to exist that does not do any border array computation at all.

\section*{Declarations}

\subsection*{Funding} Not Applicable.

\subsection*{Conflicts of interest/Competing interests} None declared.

\subsection*{Availability of data and material} Not Applicable.

\subsection*{Code availability} Not Applicable.

\bibliographystyle{unsrt}
\bibliography{bibliography}

\end{document}